\documentclass[twocolumn,prl,amsmath,amssymb,showpacs,floatfix]{revtex4}
\usepackage{bm}
\usepackage{graphicx}
\begin{document}
\title{Electromagnetic Surface Modes at Interfaces with
Negative Refractive Index\\ make a ``Not-Quite-Perfect'' Lens.}
\author{F. D. M. Haldane}
\affiliation{Department of Physics, Princeton University,
Princeton, NJ 08544-0708}
\date{June 21, 2002}

\begin{abstract}
Interfaces between media with negative 
relative refractive index generically support propagating
electromagnetic surface polariton modes with large wavenumber.
The relation of these modes to a
recent prediction by Pendry of 
``perfect (real) image formation'' by a parallel
slab of negative-refractive-index material 
is analyzed.   The ``perfect image'' theory is found to be
incomplete without inclusion of a large-wavenumber cutoff
that derives from a necessary
wavenumber-dependence of the constitutive relations, and 
which controls the resolution of the image.

\end{abstract}


\maketitle

Long ago, Veselago \cite{veselago} noted that if an object could be viewed
though a transparent slab of thickness
$w$ of a notional
material with negative relative refractive index $n$ = $-1$, simple ray optics
shows that there is a focused image at a distance $2w$ 
in front of the object, which is
real if the distance between the object and the back surface of the slab is 
less than $w$.  
In general, as seen below, 
the condition $n$ = $-1$ can be satisfied only for light
with a frequency  belonging to 
a discrete set of  one or more 
special frequencies $\omega^*$ that characterize
an interface that supports ``negative refraction''.

Ray optics is only valid at lengthscales large compared to the wavelength
of light, but recently Pendry \cite{Pendry} reported
that if the condition
$n = -1$ can be supplemented with the condition of 
{\it perfect impedance-matching}
between  electromagnetic waves in the two media
at the special frequency $\omega^*$,
the formal solution of Maxwell's equations,
with a local effective-medium approximation for the constitutive relations,
predicts that a real
image formed by light at that frequency is ``perfect'',
in that it reproduces the features of the object at all lengthscales,
however small.  Pendry describes  this counterintuitive
prediction from Maxwell's equation as ``superlensing'';
superficially, 
his solution appears mathematically correct, but it has been
controversial, and various commentators \cite{thooft,williams,valanju,garcia} 
have looked for flaws in his reasoning.

Ruppin \cite{ruppin1} has found that, unlike conventional
refracting interfaces, interfaces
with negative refractive index  support
surface electromagnetic modes (``surface polaritons'').
In this Letter, I point out that  Pendry's ``perfect image''
result, and its limitations, can be understood from a
degeneration of these modes in the impedance-matched limit.
In the local effective-medium approximation, the surface
polaritons become degenerate and dispersionless at the
``lensing frequency'' $\omega^*$, with no upper limit
to their surface wavenumber. It is this unphysical
feature that ``explains''  the ``perfect image'' prediction,
but this does not appear to have been previously explicitly noticed, 
either by Pendry (though he hints \cite{Pendry} at a 
connection to ``well-defined surface plasmons''),
or his critics, or in Ref. \onlinecite{ruppin1}.
The dispersionless surface modes are
clearly a pathology of the
approximation which neglects any wavenumber dependence of
the constitutive relations.  In fact, as with all real matter,
the microscopic nature of the lensing medium will provide
an ``ultraviolet'' (large wavenumber) cutoff.
This cutoff is relevant for optics
{\it only} in the ``superlensing'' limit, 
when it controls the actual
resolution of  real images.

\begin{figure}
\includegraphics{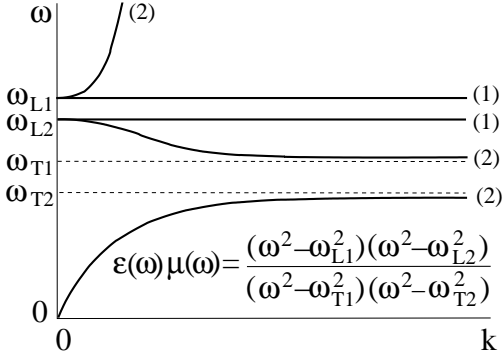}
\caption{\label{fig1}
In this model spectrum, the photon couples to 
the transverse components of two polarization modes (one electric, one
magnetic) in a material.  The propagating modes at low 
and high 
frequencies
are of the usual so-called ``right-handed'' type, 
with a positive group velocity,
but are ``left-handed'' in the frequency range
$\omega_{T1} < \omega <  \omega_{L2}$, where the group velocity 
is negative.  A surface has a negative refraction index in a frequency
range where it is an interface between bulk regions
with opposite signs of the group velocity.
}
\end{figure}

Since the direction of light rays in the
ray-optics limit is the direction of the
group velocity of the light, and the component of the wavevector parallel
to the surface is conserved during refraction, it is easy to see that 
a surface with negative relative refractive index in some
frequency range  is an interface between
media which both support propagating long-wavelength
electromagnetic waves at those frequencies, but
with {\it group velocities of opposite sign}.  
The group velocity of electromagnetic
waves must be positive in both the low-frequency and high-frequency
limits, but can be negative at intermediate frequencies, as shown
in Fig.(\ref{fig1}); in an isotropic medium, transverse modes 
with negative group velocity must become degenerate with
a finite-frequency longitudinal mode as $k \rightarrow 0$.

The Poynting vector ${\mathbf E}\times {\mathbf H}$ of a propagating electromagnetic
wave is parallel to the group velocity.  Referring to the orthogonal
triad of 
three vectors $({\mathbf E}, {\mathbf H}, {\mathbf k})$ 
that characterize propagating electromagnetic waves in an
isotropic medium,
Veselago \cite{veselago} introduced the term ``{\it left-handed media}'' 
to describe
media with negative group velocity, as 
opposed to conventional ``{\it right-handed}'' media with positive group 
velocity.  
(This terminology seems
potentially misleading, as  
the ``handedness'' it refers to derives from the 
representation of the magnetic
field as an axial vector, not intrinsic chirality
of the medium.)

Note that the model spectrum shown in 
Fig.(\ref{fig1}) will only exhibit dissipation at
the frequencies $\omega_{T1}$ and $\omega_{T2}$.   Some of the
commentators \cite{williams,garcia} have suggested that Pendry's
calculation must fail in practice because it omits absorption effects,
which, because the constitutive relations are frequency-dependent,
must be present at some frequencies because of the Kramers-Kr\"onig
relation.  However, this does {\it not} require dissipation 
in the frequency range of interest.  
In fact, if the ``left-handed'' medium can be treated
as loss-free, it must be a periodic structure, so a
microscopic lengthscale is provided by its lattice spacing.
If the material is a perfectly-periodic
``photonic crystal'', and possesses only one propagating mode 
(of each polarization) in the frequency range of interest, it can be
treated as non-dissipative to the extent that non-linearity
is negligible.

In an isotropic medium with 
frequency-dependent but {\it local} (wavenumber-independent) constitutive
relations, the spectrum of electromagnetic waves predicted by the effective
Maxwell equations is given by $c^2k^2$ = $\omega^2\epsilon(\omega)\mu(\omega)$,
where $\epsilon$ and $\mu$ are the dielectric constant and relative 
permeability.   If waves in the 
two media have opposite group velocities, there
will be a particular frequency $\omega^*$ at which the wavelengths
of propagating waves in the two media coincide:
$\lambda_1(\omega^*)$ = $\lambda_2(\omega^*)$ = $2\pi/k^*$;
the condition $n = -1$ for a flat interface
to produce a focused image is only realized at this special frequency.
The group velocity $v$ at this frequency is given by the expansion
in $ \delta\omega$ = $\omega - \omega^*$:
\begin{equation}
\omega^2\epsilon(\omega)\mu(\omega)
= c^2k^{*2}  + 2c^2k^*v^{-1}\delta\omega 
+  O(\delta\omega)^2 .
\label{exp}
\end{equation}

Let the interface be the plane $z$ = 0.
Following Ruppin \cite{ruppin1},
I look for an ``S-polarized''
interface mode
\begin{eqnarray}
B^z(x,y,z) 
&=& B^z_0 e^{\kappa_1z}e^{i(k_{\parallel}x-\omega t)},
\quad \mbox{$z \leq 0$} ,\nonumber \\
&=& B^z_0 e^{-\kappa_2z}e^{i(k_{\parallel}x-\omega t)},
\quad \mbox{$z \geq 0$} ,
\end{eqnarray}
where  $\kappa_1$ and $\kappa_2 $ are both positive;
$B^z$ couples to $H^x$ and $E^y$, and all  are continuous
at the interface:
$k_{\parallel}E^y$ = $\omega B^z$, and $k_{\parallel}\mu_0H^x$ = 
$i\alpha B^z$, where \cite{ruppin1}
\begin{eqnarray}
\alpha &=& \frac{\kappa_1}{\mu_1} = -\frac{\kappa_2}{\mu_2}.
\label{max1}
\end{eqnarray}
This  only has a solution when $\mu_1/\mu_2$ is negative.
Since by assumption
$\epsilon_1\mu_1$ and $\epsilon_2\mu_2$ are both positive,  
$\epsilon_2/\epsilon_1$ is also negative, and the interface has negative
refractive index.  The source of these fields is an oscillating 
transverse surface
polarization current
\begin{eqnarray}
J^y(x,y) &=& J^y_0e^{i(k_{\parallel}x-\omega t)}, \\
 k_{\parallel}\mu_0J^y_0 
&=& 
i\alpha(\mu_1-\mu_2)B^z_0 .
\end{eqnarray}

The condition giving the
frequency of the mode comes from combining (\ref{max1}) with
\begin{eqnarray}
c^2\kappa_1^2 &=&
c^2k_{\parallel}^2 - \omega^2\epsilon_1\mu_1 , \nonumber \\ 
c^2\kappa_2^2 &=&
c^2k_{\parallel}^2 - \omega^2\epsilon_2\mu_2 .
\end{eqnarray}
For frequencies close to $\omega^*$, the expansion (\ref{exp}) 
can be used: for small positive
$\delta k_{\parallel} $  = $k_{\parallel} -k^*$, 
$\delta \omega/\delta k_{\parallel}$ $\rightarrow$ $v$,
as $\delta k_{\parallel}$ $\rightarrow$ $0^+$, where
\begin{equation}
v  = \frac{v_1v_2(\kappa_1^2 - \kappa_2^2)}{(v_1\kappa_1^2 - v_2\kappa_2^2)},
\end{equation} 
where $v_1$ and $v_2$ are the group velocities in the two media at
frequency $\omega^*$, which have opposite signs.

This result is easy to understand: the group velocity of the surface mode is
a weighted average of the
bulk group velocities of the two media, with a larger weight given to
the velocity in the medium with smaller $\kappa$, into which the fields
penetrate deeper.
In the special case $\mu_1$ = $-\mu_2$,
$\kappa_1$ = $\kappa_2$, the competition between the two
media is exactly balanced, and $v$ vanishes; in this limit
the predicted frequency
of the mode becomes perfectly dispersionless with 
$\omega(k_{\parallel})$ =  $\omega^*$ for
all $k_{\parallel}$ $>$ $k^*$.  This corresponds to 
perfect impedance matching at frequency $\omega^*$:
$\mu_1(\omega^*)/\epsilon_1(\omega^*)$ =
$\mu_2(\omega^*)/\epsilon_2(\omega^*)$, when no reflection
of incident propagating waves with $k_{\parallel}$ $ <$ $ k^*$
will occur.

There is a second (``P-polarized'') mode, deriving from longitudinal surface 
polarization currents:
\begin{eqnarray}
J^x(x,y) &=& J^x_0e^{i(k_{\parallel}x-\omega t)}, \\
D^z(x,y,z) 
&=& D^z_0 e^{\kappa'_1z}e^{i(k_{\parallel}x-\omega t)},
\quad \mbox{$z \leq 0$} ,\nonumber \\
&=& D^z_0 e^{-\kappa'_2z}e^{i(k_{\parallel}x-\omega t)},
\quad \mbox{$z \geq 0$} ,
\end{eqnarray}
where
$\omega(\mu_2-\mu_1)D^z_0$
=  $ k_{\parallel}J^x_0$ ,
$k_{\parallel}H^y$ = $-\omega D^z$, and   
$k_{\parallel}\epsilon_0E^x$ = $i\alpha' D^z$, with \cite{ruppin1}
\begin{equation}
\alpha' = \frac{\kappa'_1}{\epsilon_1} = -\frac{\kappa'_2}{\epsilon_2}.
\end{equation}
Because 
$\epsilon_1(\omega^*)/\epsilon_2(\omega^*)$ = 
$\mu_2(\omega^*)/\mu_1(\omega^*)$, when one 
of these interface modes has positive group velocity along the interface,
the other group velocity is negative.   The spectrum is schematically
depicted in Fig.({\ref{fig2}).
\begin{figure}
\includegraphics{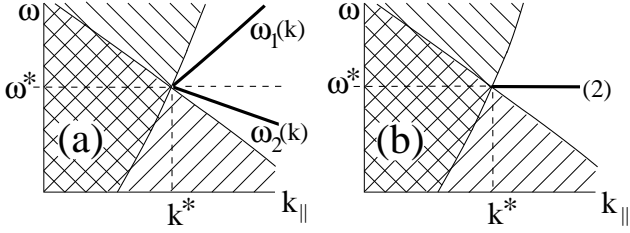}
\caption{\label{fig2}
(a): 
Generic spectrum of electromagnetic  modes 
that propagate along a negative-refractive-index
interface 
with surface wavenumber $k_{\parallel}$
$>$ $k^*$: there are two surface modes,
respectively with  positive and negative group velocity;
the shaded regions of the $(\omega,k_{\parallel})$ plane
indicate where either one or both of the media
supports propagating bulk modes, and $\omega^*$ is 
the special frequency at which waves have the same wavelength
$\lambda^* = 2\pi/k^*$ in both media.
(b): The degenerate spectrum predicted by the local
Maxwell equations in the impedance-matched 
limit where ``perfect lens'' behavior has been
predicted: the two surface modes become degenerate
at the frequency $\omega^*$ for all $k_{\parallel}$
$>$ $k^*$.
Such exactly-dispersionless degenerate
surface modes are an  
artifact of the approximation which neglects  
wavenumber-dependence of the constitutive
relations of the ``left-handed'' medium.
}
\end{figure}

It is now instructive to examine the coupling between the two sets of modes
on the opposite faces of a slab of width $d$ of a medium 2
embedded in medium 1.
The consistency condition (\ref{max1}) for the $(B^z,H^x,E^y)$ slab 
polariton modes
becomes \cite{ruppin2}
\begin{eqnarray}
\frac{\kappa_1}{\mu_1} 
&=& -\frac{\kappa_2}{\mu_2} \left( \tanh(\kappa_2d/2)\right )^{\pm 1},
\label{max2}
\end{eqnarray}
where the $\pm$  distinguishes the modes which have
symmetric $(+)$ and antisymmetric $(-)$
flux $B^z(z)$.
When $(k_{\parallel}-k^*)d \gg 1$, 
$\kappa_2d$ is very large,
and the splitting between the symmetric and antisymmetric
combinations of the modes on the two faces is very small,
proportional to $\exp (- k_{\parallel}d)$.
In the other limit, they become the $n$ = 0 and $n$ = 1 
bands of modes confined to the slab, that emerge from
the edge of the continuum of exterior propagating modes with initial
group velocity $v_1$ at $k_{\parallel}$ $\simeq$ $k^* - n^2\pi^2/2k^*d^2$,
$n \ge 0$ (the $n \ge 2$ bands remain within the region of the
$(\omega, k_{\parallel})$-plane where the slab medium has propagating modes). 
Similar considerations apply to 
the $(D^z,E^x,H^y)$ modes, but
with $\mu$ replaced by $\epsilon$.
The predicted spectrum is shown schematically in Fig.(\ref{fig3}).
There is a single band-crossing of the
surface modes (allowed because ``S'' and ``P'' polarizations
do not mix)
exactly at the frequency
$\omega^*$, 
when $\kappa_1$ = $\kappa_2$ = $\kappa_0$,
at a surface wavenumber $k_0 > k^*$  given by:
\begin{equation} 
k_0^2 = k^{*2} + \kappa_0^2 , \quad
e^{-\kappa_0d} 
= \gamma^2 < 1,
\end{equation}
where $\gamma$ is the ratio 
\begin{equation}
\gamma  =
\left ( \frac{\mu_2(\omega^*) + \mu_1(\omega^*)}
{\mu_2(\omega^*) - \mu_1(\omega^*)} \right ) =
 \left ( \frac{\epsilon_1(\omega^*) + \epsilon_2(\omega^*)}
{\epsilon_1(\omega^*) - \epsilon_2(\omega^*)} \right )
\end{equation}
(the two expressions for $\gamma$ are equivalent because
$\epsilon_1\mu_1$ = $\epsilon_2\mu_2$
at the frequency $\omega^*$).  
Note that 
$k_0$ $\rightarrow$ $\infty$
in the impedance-matched
``perfect lens'' limit 
where $\gamma$ = 0.

\begin{figure}
\includegraphics{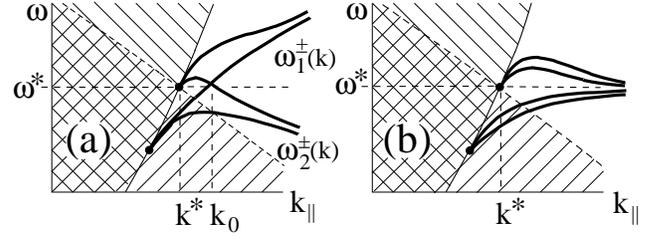}
\caption{\label{fig3}
(a): 
Generic spectrum of the coupled
electromagnetic surface modes 
of a flat slab of ``left-handed'' 
medium, with finite thickness $d$, 
embedded in a standard ``right-handed'' medium, calculated assuming
local (wavenumber-independent) constitutive relations.  The 
splitting between the even and odd combinations
of the surface modes becomes exponentially small for
$k_{\parallel}$ large. 
An allowed band crossing
occurs exactly at the  frequency $\omega^*$,
at $k_{\parallel}$ = $k_0$ (see text).
(b): Predicted spectrum
in the ``perfect lens'' limit (perfect impedance matching).
The band-crossing point $k_0$ recedes to $k_{\parallel} =  \infty$, and 
for large $k_{\parallel}$
the surface mode frequencies differ from 
the ``perfect lens'' frequency $\omega^*$ by
exponentially small splittings proportional to
$\exp (-k_{\parallel}d)$.
}
\end{figure}

I now examine the
solution of Maxwell's equations for the 
steady-state radiation field of an object
illuminated with radiation at the special frequency $\omega^*$.
The sources of the
radiation field are the oscillating currents in the object that are excited
by the illuminating field.  I will assume the source current distribution
is restricted to the region $z \le 0$, and that
the object is viewed though a slab of
``left-handed'' medium that is present in the region $0 < z_1 < z < z_2$,
where $z_2-z_1$ = $ d > 0$.
The source can be resolved  into transverse 
Fourier components ${\mathbf k}_{\parallel}$
in the $x$ and $y$
coordinates, and the radiation field of each
Fourier component computed  separately.  Consider a Fourier
component with $k_{\parallel}^2 -  k^{*2}$ =
$\kappa^2 > 0$, which produces an evanescent
field in the region $z > 0$:
\begin{eqnarray}
B^z &=& B_0e^{i(k_{\parallel}x-\omega^*t)}F(z,\gamma) , \nonumber\\
k_{\parallel}E^y &=& \omega^*B^z,\quad
-ik_{\parallel}\mu_0H^x = \mu^{-1}\frac{\partial B^z}{\partial z} , \\
D^z &=& D_0e^{i(k_{\parallel}x-\omega^*t)}F(z,-\gamma) , \nonumber\\
k_{\parallel}H^y &=&  -\omega^*D^z,\quad
-ik_{\parallel}\epsilon_0E^x = \epsilon^{-1}\frac{\partial D^z}{\partial z} , 
\end{eqnarray}
where (using continuity of $H^x$ and $E^x$ at the interfaces)
\begin{eqnarray}
&&F(z,\gamma) = e^{-\kappa z} + \alpha e^{-\kappa |z-z_1|} + 
\beta e^{-\kappa |z-z_2|} , \\
&&
\alpha =  \left ( \frac{\gamma + e^{-2\kappa d}} 
{\gamma^2 - e^{-2\kappa d} } \right )e^{-\kappa z_1}, \\
&&
\beta = - \left ( \frac{\gamma + 1} 
{\gamma^2 - e^{-2\kappa d} } \right )e^{-\kappa z_2}.
\end{eqnarray}
When presented in this form,
the solution of Maxwell's equations
has a simple physical interpretation: $F(z,\gamma)$ is the
sum of the evanescent radiation field of the object, driven 
directly by the illumination, plus the
radiation fields of the induced surface polarization currents of the
slab, driven by the radiation field of the object.
The amplitudes of the two surface currents diverge
when $k_{\parallel}$ = $k_0$ ($\exp -2\kappa d = \gamma^2$), when the
driving frequency $\omega^*$ is exactly resonant with the coupled
surface modes.  This resonance recedes to $k_{\parallel}$
= $\infty$ as the impedances are tuned to the ``perfect lens''
limit $\gamma = 0$.

When $\gamma$ = 0,
the field of the small surface
polarization current on the left interface 
at $z_1$ exactly cancels the field of the object
for $z > z_1$, while for $z < z_1$, it cancels the field of the second 
polarization  current on the right interface at $z_2$.   
The field observed for $z>z_1$ is then just  the radiation field
of the surface polarization current at $z_2$. 
If $z_1 < d$, this current has an
amplitude that, as $k_{\parallel} \rightarrow \infty$, 
{\it grows exponentially} $\propto \exp k_{\parallel} (d -z_1)$
relative to the strength of the source radiation field.
Since the object was taken to be to the left of the plane
$z=0$, the condition $z_1 < d$ is precisely the condition that 
a real image of some part of the object can be formed.

The predicted exponentially-large amplification
in the large-wavenumber limit limit occurs because
{\it the difference between the coupled surface mode frequencies
and the driving frequency $\omega^*$  becomes exponentially small}
as $k_{\parallel}$ $\rightarrow $ $ \infty$ (see Fig.(\ref{fig3}b).
It is a pathology of Pendry's solution that was 
noted in Ref. \onlinecite{garcia}; however,
while those authors recognized that, for
$|z-z_2| < d-z_1$, this exponential amplification leads to an
 ``ultra-violet'' (large wavenumber)
divergence of the expression for the predicted radiation field 
of a point object at $z=0$, they drew
the incorrect conclusion that this divergence implied that the evanescent
radiation field of such a point object
cannot penetrate the slab.   In fact, the divergence
will be always controlled by the short-distance cutoff provided by the physical
nature of the interface; this may either be the maximum wavenumber
of a surface polarization current ({\it e.g.}, a surface Brillouin zone
boundary), or a wavenumber at which $k$-dependent constitutive relations
move the surface mode frequencies away from near-resonance with the 
driving frequency
$\omega^*$.

Pendry's solution is only strictly valid without a cutoff in the
case $z_1 > d$, but remains non-singular in the marginal case
$z_1 = d$, where the image of a source
at $z=0$ is neither real nor virtual, but is exactly  {\it on}
the right surface $z=z_2$ of the slab.  In this limit,
there is an interesting interpretation of Pendry's formal result:
the replacement $\epsilon \rightarrow -\epsilon$,
$\mu \rightarrow -\mu$ is equivalent to the transformation
$({\mathbf B}, {\mathbf D}, {\mathbf E}, {\mathbf H})$ $\rightarrow  $
$(-{\mathbf B}, -{\mathbf D}, {\mathbf E}, {\mathbf H})$, which in turn
is equivalent to
time-reversal of the source-free Maxwell equations.   The predicted 
``perfect image'' formed by monochromatic
radiation with frequency $\omega^*$ is then analogous to the
``spin-echo'' after a $\pi$-pulse in magnetic resonance: the
radiation field from a source at $z=0$  first 
propagates a distance $d$ along the $z$-axis
in the normal  medium, with dispersion of its 
${\mathbf k}_{\parallel}$-Fourier-components, 
then this dispersion is exactly reversed
by subsequent  propagation through an equal thickness  $d$ of 
the ``time-reversing''
medium to refocus the field  at  the ``perfect image'' point.  However, this
intriguing interpretation is spoiled by the 
inequivalence between a physical ``left-handed''
medium and an idealized 
``time-reversed vacuum'': the equivalence
is  only valid at a single frequency
$\omega^*$, and as $k\rightarrow 0$
the frequency of ``left-handed light''  approaches a finite
value $\omega_L$ (see Fig.(\ref{fig1})); also, the medium
will provide a large-wavenumber cutoff.

In summary, I have shown Pendry's \cite{Pendry} 
controversial  ``superlens'' theory is incomplete,
in that it fails to explicitly include a large-momentum
cutoff, which is needed to regularize the theory
when it describes a real image.    The high Fourier components
of the image are produced by the near-fields of surface polariton
modes \cite{ruppin1,ruppin2} which in the ``superlens'' limit
are exponentially close to resonance at the special lensing frequency
$\omega^*$.
The microscopic structure of the surface of
the  lensing medium will determine this cutoff,
which is unspecified in a local-effective medium approximation,
and will limit the resolution of the image.   While
``not-quite-perfect'', it seems that this resolution can in 
principle be engineered to be significantly smaller than the
wavelength of the illuminating radiation, without violating
any fundamental physical principles.

This work is supported in part by NSF MRSEC 
DMR-98094983 at the Princeton Center
for Complex Materials.

\end{document}